\documentclass[a4paper,10pt,twocolumn,aps,prl,nolongbibliography,superscriptaddress,showpacs,showkeys,amsmath,amssymb,floatfix,nofootinbib]{revtex4-1}
\usepackage{lmodern}

\usepackage[T1]{fontenc}
\usepackage[utf8]{inputenc}
\usepackage{color}
\usepackage[unicode=true,pdfusetitle,
 bookmarks=true,bookmarksnumbered=true,bookmarksopen=true,bookmarksopenlevel=1,
 breaklinks=false,pdfborder={0 0 0},backref=false,colorlinks=true]
 {hyperref}
\hypersetup{
 citecolor=blue,filecolor=blue,linkcolor=blue,urlcolor=blue}
\makeatletter

\pdfpageheight\paperheight
\pdfpagewidth\paperwidth

 \@ifundefined{textcolor}{}
 {%
   \definecolor{BLACK}{gray}{0}
   \definecolor{WHITE}{gray}{1}
   \definecolor{RED}{rgb}{1,0,0}
   \definecolor{green}{rgb}{0,0.7,0}
   \definecolor{BLUE}{rgb}{0,0,1}
   \definecolor{CYAN}{cmyk}{1,0,0,0}
   \definecolor{MAGENTA}{cmyk}{0,1,0,0}
   \definecolor{YELLOW}{cmyk}{0,0,1,0}
 }

\renewcommand{\[}{\begin{equation}}
\renewcommand{\]}{\end{equation}}

\usepackage{array}
\makeatother

\begin{document}

\title{The fate of large-scale structure in modified gravity after GW170817 and GRB170817A}

\author{Luca Amendola}
\affiliation{Institut für Theoretische Physik, Ruprecht-Karls-Universität Heidelberg,
Philosophenweg 16, 69120 Heidelberg, Germany}

\author{Martin Kunz}
\affiliation{Départment de Physique Théorique and Center for Astroparticle Physics,
Université de Genève, Quai E. Ansermet 24, CH-1211 Genève 4, Switzerland}

\author{Ippocratis D.~Saltas}
\author{Ignacy Sawicki}
\affiliation{CEICO, Fyzikální ústav Akademie v\v{e}d \v{C}R, Na Slovance 2, 182 21 Praha 8, Czechia}

\begin{abstract}
The coincident detection of gravitational waves (GW) and a gamma-ray burst from a merger of neutron stars has placed an extremely stringent bound on the speed of GW. We showed previously that the presence of gravitational slip ($\eta$) in cosmology is intimately tied to modifications of GW propagation. This new constraint implies that the only remaining viable source of gravitational slip is a conformal coupling to gravity in scalar-tensor theories, while viable vector-tensor theories cannot now generate gravitational slip at all. We discuss structure formation in the remaining viable models, demonstrating that (i) the dark-matter growth rate must now be at least as fast as in GR, with the possible exception of the beyond Horndeski model. (ii) If there is any scale-dependence at all in the slip parameter, it is such that it takes the GR value at large scales.  We show a consistency relation which must be violated if gravity is modified.
\end{abstract}
\maketitle

\section{Introduction}
The recent detection of a gravitational-wave and electromagnetic signal from the merging of two neutron stars \cite{TheLIGOScientific:2017qsa} (events GW170817 and GRB 170817A, hereinafter ``the Event'') provides not only an exciting discovery, but also strongly challenges the observational viability of large classes of gravitational theories for the late Universe as anticipated in \cite{Amendola:2012ky} and discussed e.g.\ in \cite{Creminelli:2017sry,Ezquiaga:2017ekz,Sakstein:2017xjx,Baker:2017hug}. Theories aiming to describe the late-time acceleration of the Universe introduce novel, non-trivial interactions between the spacetime metric and new extra degrees of freedom such as scalar (Horndeski \cite{Horndeski:1974wa,Deffayet:2011gz}, vector (Einstein-Aether \cite{Jacobson:2000xp}, generalized Proca \cite{Heisenberg:2014rta,Tasinato:2014eka}) and tensor (bi-gravity) fields. 

In \cite{Saltas:2014dha,Sawicki:2016klv} it was shown for the first time that {\it a precise link between tensor and scalar fluctuations in cosmology exists}: a modification to the gravitational-wave propagation at any scale implies the existence of gravitational slip ($\eta\neq1$) for large-scale scalar fluctuations, with both modifications driven by exactly the same theory-space parameters of the gravitational model. The gravitational slip leaves a particular and observable imprint on the formation of structures in the universe, which can be used to constrain models of late-time acceleration \cite{Amendola:2012ys,Amendola:2012ky,Saltas:2010tt,Motta:2013cwa}. 

The Event has allowed LIGO to measure the speed of GWs with precision $\left|c_\text{T}/c-1\right| \leq 1\times10^{-15}$ \cite{Monitor:2017mdv}. For all intents and purposes, from the point of view of cosmology, the speed of GWs at the present time is now  known to be that of light. By the above argument, the range of possible scenarios for structure formation is also narrowed.  In this Letter, we ask the question: {\it What are the key implications of the Event for the phenomenology of large-scale structure?}

We will discuss below the implications for each class of models of acceleration featuring one extra degree of freedom in turn, introducing here only the essential notation and definitions. Considering the line element of scalar fluctuations in Newtonian gauge, $\mathrm{d}s^2 = -(1 + 2\Psi(\boldsymbol{x},t))\mathrm{d}t^2 + a^2(t)(1 -2\Phi(\boldsymbol{x},t))\mathrm{d}\boldsymbol{x}^2$, we define the gravitational slip as $\eta \equiv \Phi/\Psi\neq 1$ and the respective effective Newton's constants in momentum space $Y \equiv -2 k^2 \Psi/(a^2 \rho_m \delta_m)$ and $Z=\eta Y$, where $\delta_m$ is the comoving matter density contrast. Our working {\it definition of modified gravity} is the one introduced in \cite{Saltas:2014dha}, i.e.\ any gravitational model modifying the linear propagation of tensor modes compared to GR, i.e.\ which by \cite{Saltas:2014dha,Sawicki:2016klv} produces gravitational slip from perfect-fluid sources.

In this paper, we start with the remaining viable {\it scalar-tensor models} of gravity. With the new constraint, they can have at most a conformal coupling to curvature \cite{Creminelli:2017sry,Ezquiaga:2017ekz,Sakstein:2017xjx,Baker:2017hug}, which is now  the only admissible cause of gravitational slip from perfect-fluid matter. We demonstrate that, if slip is generated at all, it either has no scale dependence at linear scales, or it disappears at large scales, with the theory of gravity returning to $\eta=1$ there. We show that the growth rate must be higher than GR for all models with the possible exception of the single remaining class of `beyond Horndeski' theories.

Then we show that the remaining viable {\it vector-tensor theories} cannot generate slip at all and in no remaining viable such model can the growth rate be lower than in GR.

The Event has not placed new constraints on {\it theories of massive gravity} \cite{Baker:2017hug}. We do not study these further, since there is no single model of massive (bi)-gravity which could account for the whole of cosmological history without some sort of pathology \cite{D'Amico:2011jj,Konnig:2015lfa,Lagos:2014lca,Cusin:2014psa} unless it has the same predictions as $\Lambda$CDM \cite{Akrami:2015qga}.

\section{Dark energy phenomenology}

\subsection{Scalar-tensor theories: Horndeski}
Horndeski theories are the most general scalar-field theories which have equations of motion with no higher than second derivatives \cite{Horndeski:1974wa} and where all matter is universally coupled to gravity.  They include as a subset the archetypal modifications of gravity such as $f(R)$ and Brans-Dicke theories, as well as galileons \cite{Nicolis:2008in,Deffayet:2009wt}.The popular dark energy models of quintessense \cite{Ratra:1987rm,Wetterich:1987fm} and k-essence \cite{ArmendarizPicon:2000ah} are also subclasses of Horndeski. At the level of the action, they are described by four, in principle arbitrary, coupling functions $G_{2,3,4,5}(X,\phi)$ where $X \equiv -(\partial \phi)^2/2$. At the same time, it has been shown \cite{Bellini:2014fua} that the dynamics of  linear fluctuations on the cosmological background are completely characterized by four (time-dependent) functions: the kineticity $\alpha_\mathrm{K}(t)$, related to the Jeans scale for the scalar, the braiding $\alpha_\mathrm{B}(t)$, measuring the degree of kinetic mixing between the scalar and the metric, the running of Planck mass $\alpha_\text{M}(t)$, and the tensor speed excess $\alpha_\mathrm{T}(t)$. In particular, $\alpha_\text{T} \equiv c_\mathrm{T}^2 - 1$ measures the departure of the GW’s speed from that of light and is the parameter constrained to be effectively zero by the Event \cite{TheLIGOScientific:2017qsa}. Requiring that this not be achieved by a severe tuning of the parameters implies that the scalar can at most be  coupled conformally to curvature, i.e.\ $G_5=0$ and $G_4=G_4(\phi)$ \cite{Creminelli:2017sry,Ezquiaga:2017ekz,Baker:2017hug}. This means that the most general Horndeski model still allowed has the Lagrangian
\begin{equation}
	\mathcal{L} = \frac{f(\phi)}{2}R + K(X,\phi) - G(X,\phi)\Box\phi \label{eq:KGBf}\,,
\end{equation}
i.e.\ belongs to the class of kinetic gravity braiding (KGB \cite{Deffayet:2010qz}) augmented by a conformal coupling to gravity. Setting $G=G(\phi)$ reduces these models to conformally coupled k-essence \cite{Babichev:2009ee,Sawicki:2012re}, while setting $K=V(\phi),\, G=0$ is equivalent to $f(R)$ gravity. Setting $f=\text{const}$, means gravity is no longer modified, but the model is nonetheless capable of accelerating the expansion of the Universe without a cosmological constant \cite{Deffayet:2010qz}. 

The class of models \eqref{eq:KGBf} is significantly more restricted than full Horndeski, thus we can make precise predictions for large-scale structure formation. In particular, using the definitions and results of \cite{Bellini:2014fua}, we have that in the small-scale limit, yet still linear regime within the quasi-static approximation (i.e. $k\to\infty$),%
\footnote{In ref.~\cite{Sawicki:2016klv}, we showed that under extremely fine-tuned choices of a Horndeski action, it would in principle be possible to preserve a configuration with $Y=Z$ everywhere under time evolution, thus dynamically shielding the modification of gravity. The measurement of $\alpha_\text{T}=0$ means \emph{none of those models are still viable}, thus if the coupling to gravity is not minimal, the configuration of the fluctuations will reflect it.}
\begin{equation}
	Y_\infty = 1 + \frac {(\varkappa+\alpha_M)^2}{2N}\,,\quad
	Z_\infty = 1 + \frac{\varkappa^2-\alpha_M^2}{2N}\, , \label{eq:YZ}
\end{equation}
where $\alpha_M=\dot\phi f_{,\phi}/Hf$ is the rate of evolution of the effective Planck mass. The function $\varkappa\equiv \alpha_B+\alpha_M$ is the part of the braiding produced by the the term $G(X)$, it is zero in Brans-Dicke, k-essence and $f(R)$ models. $N$ is the numerator of the sound speed of the scalar and must be positive definite (the denominator of the sound speed is positive as a result of the no-ghost condition),
\begin{align}
	 N\equiv& -(2+\alpha_M)\dot{H}/H^2 + 3\Omega_m/f +\alpha_M(2+\alpha_M) \\  
     &-\alpha_M' +\varkappa(2-\varkappa)/2-\varkappa \dot{H}/H^2 +\varkappa' \, . \notag
\end{align}
We can combine these results to obtain
\begin{equation}
\eta_\infty -1 = -\frac{2\alpha_M(\varkappa+\alpha_M)}{2N+(\varkappa + \alpha_M)^2} \,.
\end{equation}
{\it We can thus make general statements about the properties of gravity for the remaining scalar--tensor theories at small linear scales:}
\begin{itemize}
\item The effective Newton's constant for non-relativistic matter, $Y_\infty\geq1$, so in the remaining Horndeski models, matter cannot cluster slower than in GR given the same background and the same $\Omega_m$ \cite{Piazza:2013pua}
\item The effective Newton's constant for the lensing potential, $\Sigma\equiv(Y+Z)/2$ is different from unity whenever the KGB term is present, $\varkappa\neq 0,\alpha_M$.
\item The gravitational slip parameter $\eta_\infty\equiv Z_\infty/Y_\infty$ can be both larger or smaller than unity.
\end{itemize}
If the KGB term is not present, $\Sigma_\infty=1$ and $\eta_\infty\leq1$, thus a violation of either of these conditions can be interpreted as a detection of the presence of kinetic gravity braiding. When gravity is minimally coupled $f=\text{const}$, $\alpha_M=0$. This gives $Y_\infty=Z_\infty>1$ and $\eta=1$ at all scales; in this case, if the modification is large enough, the sign of the cross-correlation of the galaxies and the integrated Sachs-Wolfe effect can reverse, which is quite strongly disfavored by data \cite{Barreira:2014jha,Renk:2017rzu}.

For a generic  lagrangian \eqref{eq:KGBf}, the fluctuations of the scalar field will have a mass $M$ (see e.g.\ Ref.\ \cite{DeFelice:2011hq,Amendola:2013qna} for the expression).  If $M\lesssim H$, then the expressions  \eqref{eq:YZ} are valid at all linear scales inside the sound horizon of the scalar, $c_\text{s}k\gg H$ \cite{Sawicki:2015zya}. If $H \ll M \ll k_\text{NL}$, where $k_\text{NL}$ is a scale associated with non-linearities either in the dark matter or screening of gravity, then a transition will occur and 
\[
\eta\rightarrow 1\,,\qquad k<M\,,
\]
recovering the GR result for slip. Note that theories modifying $c_\text{T}$ do not recover this GR result at large scales. Clearly, for large masses $M\gg k_\text{NL}$, no modification of gravity is observed in linear structure formation at all.

\subsection{Scalar-Tensor Theories: Beyond Horndeski}

`Beyond Horndeski' models extend the Horndeski action by allowing for higher-order equations of motion. As a result of having degenerate Hamiltonians they nonetheless propagate no extra d.o.f.\ beyond  one scalar and two tensors \cite{Zumalacarregui:2013pma,Gleyzes:2014dya,Deffayet:2015qwa,Crisostomi:2016tcp}. There exists exactly one choice of term additional to Lagrangian \eqref{eq:KGBf} which does not affect the speed of GWs: combining the quartic Horndeski term with the quartic beyond-Horndeski term and setting $F_{4,X}=-G_{4,X}$ \cite{Creminelli:2017sry,Sakstein:2017xjx,Baker:2017hug}. This is the case, since the overall coupling of the scalar to curvature remains  conformal, but with a function of $X$ instead of just $\phi$, giving $\eta\neq1$.  One may argue that this particular choice of $F_4$ is not necessarily a tuning, since in flat space the two Lagrangian terms  are related through a total derivative and so their curved-space corrections should be suppress by the Planck mass. The new physics in large-scale structure can be described by a new parameter $\alpha_\text{H}$. General beyond-Horndeski models can have $Y_\infty < 1$, for a large-enough $\alpha_\text{H}$ and thus reduce the growth rate w.r.t. GR on the same background \cite{DAmico:2016ntq}. In the remaining viable models, however, $\alpha_\text{H}$ is related to $\alpha_\text{M}$, the rate at which the Planck mass evolves, which is constrained by observations \cite{Hart:2017ndk}.   We leave the detailed analysis of whether these constraints can be evaded sufficiently to reduce growth rates  for future work.       

\subsection{Vector-Tensor Theories}
There are two classes of modifications of gravity featuring vectors: (i) the Einstein-Aether (EA) model \cite{Jacobson:2000xp} and its generalization \cite{Zlosnik:2006sb}, and (ii) the generalized Proca theory \cite{Tasinato:2014eka,Heisenberg:2014rta} along with its `beyond generalized Proca' generalized version, similar to the `beyond Horndeski' \cite{Heisenberg:2016eld}. Class (i) is also closely related to the low-energy limit of Ho\v{r}ava-Lifshitz theories \cite{Blas:2010hb}.

In (generalized) EA models, one removes a would-be ghost by constraining the magnitude of the vector field using a Lagrange multiplier. As we discussed in our first paper on this topic \cite{Saltas:2014dha}, the same term in the action gives a source for gravitational slip from perfect-fluid matter in cosmology, and changes the speed of propagation of GWs: in the language of ref.~\cite{Lim:2004js}, $c_\text{shear}=\beta_1+\beta_3$; the result for generalized EA theories is the same \cite{Zuntz:2010jp}. As a consequence of the Event, the constraint $c_\text{shear}=0$ applies to all these models \cite{Baker:2017hug}. This means that (generalized) Einstein-Aether models cannot anymore produce gravitational slip from perfect-fluid matter, $\eta=1$ at all scales. Moreover, the effective Newtons' constant in EA  is now $Y_\infty=(1+3\beta_2)/(1-\beta_1)$ with both $\beta_{1,2}>0$ for stability, i.e.\ $Y_\infty\geq1$ and growth rates on the same background must be higher than in GR. In generalized EA, the expression has the same form, with the replacement $\beta_i\rightarrow \mathcal{F}_\mathcal{K}\beta_i$ \cite{Zlosnik:2007bu} and the same conclusions can be reached.

Generalized Proca theories, on the other hand, are the most general theories with second-order equations of motion which propagate a massive vector in addition to the graviton, with spurious degrees of freedom eliminated through a non-linearly realized Abelian gauge symmetry. The general Lagrangian is described by five functions of the vectors' magnitude $X\equiv-A_\mu A^\mu /2$, $G_{3,4,5,6}$ and $g_5$ and one function $G_2$ of $X$, the vector kinetic term $F^{\mu\nu}F_{\mu\nu}$ and the magnitude of $F^{\mu\nu}A_\nu$, resulting in a structure similar to Horndeski models (see ref.~\cite{Heisenberg:2014rta,Allys:2015sht,Jimenez:2016isa,Allys:2016jaq}). There are two branches of solutions in these theories, only one of which is dynamical. In the dynamical branch, the vector is explicitly coupled to curvature through the functions $G_4$ and $G_5$, which modifies the speed of GWs. Requiring that GWs propagate at the speed of light leads to the constraint that $G_5=0$ and $G_4=\text{const}$. $2G_4$ can be then identified with the effective Planck mass squared and is not a free parameter of the theory \cite{Baker:2017hug}.

We have re-analyzed the implications of this constraint on structure formation in these models. Ref.\ \cite{DeFelice:2016uil} showed that the general dynamics of scalar perturbations on a cosmological background depend on seven functions of time, $w_i$, which similar to the case of Horndeski theories, depend on the free functions appearing in the action. Taking $G_5=0$ and $2G_4=1$, this set reduces to just two: $w_2$ and $w_3<0$, where the sign is fixed by requiring that the propagating vectors not be ghosts. For all these models, there is no gravitational slip, $\eta=1$, and the short-scale gravitational constant reduces to
\begin{equation}
	Y_\infty=1+\frac{-w_3w_2^2}{N}\,, \qquad N\equiv 2\mu_2/\phi +w_3w_2^2 \,,
 \end{equation}
where $\mu_2$ is a function of the $w_i$ defined in \cite{DeFelice:2016uil}. For $N>0$, $Z=Y\geq1$ and gravity is stronger that in $\Lambda$CDM. This is \emph{always} the case for a model with effective dark energy (DE) equation of state parameter $w_{\rm DE}\leq-1$, since the sound speed of the scalar is only positive when $N-4w_{3}\left(2H+w_{2}\right)^{2}\rho_{\rm DE}(1+w_{\rm DE})>0$. In addition, in the future de Sitter attractor common to these theories, we necessarily have $N>0$. Changing the sign of $N$ at an earlier time would be in principle possible for a dark energy with $w_{\rm DE}>-1$, but that would lead to $N=0$ and thus a divergent effective Newton's constant at least at one moment in time. We thus conclude that $Y\geq 1$ at all times in the remaining viable generalized Proca models. 

In the same spirit as `beyond Horndeski', there exists a more general action for vectors (`beyond generalized Proca'). It is possible to write down four additional terms in the action, $f_{4,5}$ and $\tilde{f}_{5,6}$, with arbitrary functions of $X$ as coefficients \cite{Heisenberg:2016eld}. Then $f_{4,5}$ enter the speed of GWs and must be set to zero as a result of the Event. It was shown in ref.~\cite{Heisenberg:2016eld} that, on FRW, the scalar perturbations depend on one more function of time, $w_8$, but, again, requiring $c_T=1$ reduces both the equations of the background and the quadratic action for perturbations to be exactly the same as in the case of the generalized Proca, i.e.\ there is no new phenomenology allowed in the scalar sector of linear cosmology in this much wider set of theories.

We can thus conclude that the Event has constrained general vector-tensor theories and the low-energy limit of Ho\v{r}ava-Lifshitz theories to be unable to produce gravitational slip from dust, and the growth rate in the remaining viable models must be at least as fast as in GR for the same background and $\Omega_m$.

\section{Constraining the scalar mass}

As shown in \cite{Amendola:2012ky}, one can obtain three quantities from cosmological large-scale structure power spectra in the linear regime such that the dependence on the power spectrum shape, which cannot be known without an assumption for the theory of gravity, cancels out:
\begin{eqnarray}
P_{1} & \equiv & R/A=f/b \, ,\\
P_{2} & \equiv & L/R=\Omega_{\text{m}0}Y(1+\eta)/f \, ,\\
P_{3} & \equiv & R'/R=f+f'/f \, ,
\end{eqnarray}
where $b$ is the dark matter-galaxy bias and a prime stands for a derivative with respect to $\log a$.
The second quantity is also often called $E_G$ \cite{Zhang:2007nk}, while the third one is related to the commonly used quantity $f\sigma_8(z)$ (see \cite{Percival:2008sh}) by the following relation
\begin{equation}
P_3=\frac{(f\sigma_8(z))'}{f\sigma_8(z)} \, .
\end{equation}
Observationally, $P_3$ can be obtained by taking finite differences across redshift bins
\begin{equation}
P_3\approx -(1+z)\frac{\Delta (f\sigma_8(z))}{f\sigma_8(z)\Delta z} \, .
\end{equation}
In \cite{Motta:2013cwa}, we showed that the assumption of the weak equivalence principle for galaxies is enough to write in a general theory of modified gravity
\begin{equation}
\frac{3P_{2}(1+z)^{3}}{2E^{2}\left(P_{3}+2+\frac{E'}{E}\right)}-1=\eta\,,\label{eq:kpol}
\end{equation}
where $E(z)\equiv H(z)/H_0$ is the dimensionless Hubble function. This relation is valid for any cosmology and scale, regardless of $Y$, of bias, and of initial conditions.  We can now form a null relation which is violated whenever gravity is modified
\begin{equation}
P_{2}=4E^{2}\frac{\left(P_{3}+2+\frac{E'}{E}\right)}{3(1+z)^3} \,.
\end{equation}
Since the only remaining viable modified gravity models which generate slip are scalar-tensor, any violation of this relationship will be evidence that the effective Planck mass evolves in time. We have argued here that in the remaining parameter space, this relation will either be scale independent until the sound horizon, or will return to the GR value at $k<M$, giving a method to constrain the mass of the degree of freedom modifying gravity. 

\section{Summary and Conclusions}

The observation of coincident gamma radiation and GWs from the same source at cosmological distances by LIGO/VIRGO and Fermi/Integral has put so strong a bound on any deviation of the speed of GWs from that of light that, for the purposes of cosmology, any dynamics of modified gravity that cause a change in the speed of propagation at the present epoch must be completely irrelevant. In each class of gravitational theories beyond GR, this severely limits the viable theory space. 

We previously proved that there is a one-to-one relationship between the modification of the propagation of GWs and the sourcing of gravitational slip in the presence of perfect-fluid matter. Above new constraint in turn significantly reduces the sort of configurations/sources of slip that are still allowed, leading to strong observational consequences.

In this paper, we have shown that, in the newly restricted viable parameter space of universally coupled modified gravity theories, it is impossible to reduce the growth rate in structure formation at small, linear scales \emph{with respect to standard gravity on the same background and dark matter density}. This applies to Horndeski scalar-tensor models and any vector-tensor theories. Moreover, gravitational slip in the presence of perfect-fluid matter can only be produced by a conformal coupling in scalar-tensor models, and therefore an evolving Planck mass. As a direct observational consequence, it follows that a future detection of gravitational slip would exclude all vector-tensor and Ho\v{r}ava-Lifshitz Lorentz-violating models.  

If slip be present, in the remaining viable models, it is either constant on scales inside the sound horizon of the scalar or, for models with a sufficiently large mass, it is screened away to its GR value $\eta=1$ at scales above the Compton length. We have further shown how slip can be measured and thus this mass constrained in all remaining viable theories which generate it.

We note however that, it is in principle possible that growth rates can be reduced w.r.t.\ GR in `beyond Horndeski' models, although we leave the exact limits of this to a future study. Although we have not discussed the case of massive (bi-)gravity models, for the sake of completeness, let us state that it is also possible to choose parameters in massive bimetric theories so that $\eta\neq1$ at large scales and $Y<1$ at small linear scales, at least for some period of time, without instabilities during that time. However, these theories cannot actually provide a complete cosmological background from the Big Bang to today that does not suffer from pathologies at some point in the course of the cosmological evolution \cite{D'Amico:2011jj,Konnig:2015lfa,Lagos:2014lca,Cusin:2014psa}, apart from the limit where the theory behaves exactly like $\Lambda$CDM \cite{Akrami:2015qga}. 

To conclude, GWs have provided an extremely strong constraint on possible modifications of gravity at both large and small scales. This in turn has restricted the possible modifications to the evolution of large-scale structure in a very sharp manner, removing some of the freedom resulting from a large model space. This will only serve as to increase the power of upcoming cosmological surveys to constrain or eliminate the remaining viable models. 

\begin{acknowledgments}
\emph{Acknowledgements.}  The work of L.A.~is supported
by the DFG through TRR33 ``The  Dark Universe''. M.K.~acknowledges
funding by the Swiss National Science Foundation. I.S.~and I.D.S.~are supported by ESIF and MEYS (Project CoGraDS -- CZ.02.1.01/0.0/0.0/15\_003/0000437).
\end{acknowledgments}
\bibliographystyle{utcaps}
\bibliography{AnisoRefs}

\end{document}